\documentclass[prd,showpacs,amsmath,amssymb,superscriptaddress,preprintnumbers,nofootinbib,showkeys]{revtex4}

\begin{document}

\title{Extension of the Einstein-Dirac-axion-aether theory \\ based on an effective metric with a spinor kernel }

\author{Alexander B. Balakin}
\email{Alexander.Balakin@kpfu.ru} \affiliation{Department of
General Relativity and Gravitation, Institute of Physics, Kazan
Federal University, Kremlevskaya str. 16a, Kazan 420008, Russia}

\author{Anna O. Efremova}
\email{anna.efremova131@yandex.ru} \affiliation{Department of
General Relativity and Gravitation, Institute of Physics, Kazan
Federal University, Kremlevskaya str. 16a, Kazan 420008, Russia}

\date{\today}

\begin{abstract}
We consider an axionic extension of the Einstein-Dirac-aether theory, which is based on the spinor modification of the kinetic term of the pseudoscalar field and describes the backreaction of the spinor field on the axionic dark matter. The main instrument of this extension is the effective metric constructed using the aether velocity four-vector and a kernel, which depends on the basic spinor scalar and on the square of the basic spinor pseudoscalar. The periodic potential of the axion field includes a guiding function, which regulates the dynamics of axions and predetermines the properties of their equilibrium states. This guiding function and the kernel of the effective metric, as well, are considered to be functions  of the expansion scalar of the aether flow. The master equation for the guiding function is obtained as a consequence of the Lagrangian invariance with respect to the discrete transformations  prescribed by the axion field shift symmetry. The self-consistent set of coupled master equations is derived, which includes new source-terms in the modified master equations for the spinor, axion, vector and gravitational fields. Cosmological applications of the extended theory are considered; new exact solutions of these equations are presented for the isotropic homogeneous model of the Universe evolution. Discussion is focused on the mechanism of backreaction of the
spinor field on the axionic dark matter evolution.

\end{abstract}
\maketitle

\section{Introduction}

According to the modern version of the Universe history, basic proto-configurations of fermion and boson subsystems were close to be formed at the end of inflation stage. In order to indicate the process of creation of the corresponding scalar constituent, Damour and Esposito-Far\`ese in 1996 have introduced the special term "spontaneous scalarization" \cite{SS0}. This fruitful phenomenological idea about a spontaneous phase transition has been used in various astrophysical and cosmological contexts (see, e.g.,  \cite{SS1}-\cite{SS5}), and applied to the models with vector field (see, e.g., \cite{SV1,SV2,SV3}), tensor field \cite{ST}, spinor field \cite{SSpin0,SSpin1,BE0,BE01}, color dynamic aether \cite{CA2} and the gauge fields \cite{SGG}.

At the same time it is clear that when fermion and boson systems grow to cosmic scales, the reverse processes cannot be ignored, and we have to consider their backreacrion on the cosmodynamics. In this work we intend to analyze a model of such feedback in the framework of the Einstein-Dirac-aether-axion theory. For this purpose we consider the interaction of four physical fields. The first one is the gravity field described in the Einstein-Hilbert scheme.  Second, we consider the  unit timelike vector field $U^j$, associated with the velocity four-vector of the dynamic aether (see, e.g., \cite{J1,J2,J3} for basic definitions and references). Third, we analyze the evolution of the spinor field, which describes fermions with both: nonzero and vanishing rest masses. The four constituent of the model is the axion field, which is distinguished from the general class of the pseudoscalar fields by the requirement of specific discrete shift symmetry, and describes an axion fraction of the cosmic dark matter \cite{Ax1}-\cite{Ax8}.

The theory, which we study, belongs, in fact, to the category of vector-tensor-(pseudo)scalar-spinor modifications of gravity \cite{Odin1,Odin2}. The presence of the unit vector field $U^j$ in this theory realizes the idea of a preferred frame of reference (see, e.g., \cite{N1,N2}), and indicates the possibility of violation of the Lorentz invariance \cite{LV1}.
We restrict ourselves by the classical phenomenological version of the theory, since the quantum version of the aetheric vector field is not yet established, and there are no ideas what particles could be the carriers of the corresponding interactions.

What new theoretical details do appear in the presented work? We assume that the Lagrangian of the Einstein-aether theory and the Lagrangian of the spinor field are of the canonic form. Earlier, we have analyzed the catalogue of the possible spinor modifications of the Jacobson constitutive tensor, which enters the kinetic term of the Einstein-aether theory, and studied the model, in which the potential of the axion field contains the spinor invariants (see \cite{STFI2} and \cite{STFI1}, respectively).
In this paper the spinor modifications are introduced only into the kinetic part of the axion field Lagrangian.
We used the known mathematical instrument, which can be indicated as {\it effective aetheric metric with spinor kernel}, $G^{mn}{=}g^{mn} {+} {\cal A} U^m U^n$ (we use below the short term {\it effective metric}). Here $g^{mn}$ relates to the physical spacetime metric, $U^m$ is the aether velocity four-vector, and ${\cal A}$ is the kernel of the effective metric.

For the first time a mathematical construction of this kind was introduced by Gordon in \cite{Gordon}, $g_{*}^{mn}=g^{mn} {+} (n^2{-}1) V^m V^n$; it was called the optical metric, $V^m$ was the velocity four-vector of the medium, and ${\cal A}=(n^2{-}1)$, where $n$ was the refraction index of an arbitrary moving isotropic, transparent, nondispersive dielectric medium. The optical metric played an important role in the development of the covariant phenomenological electrodynamics of continuous media (see, e.g., \cite{Sing,MauginJMP,HehlObukhov,BZ} for details and references).
The theory of the acoustic metrics describes sound waves as quasi-particles, which move in the effective spacetime with the metric of the mentioned type (see, e.g., \cite{Volovik}). In \cite{BDZ1} the theories  of color and color-acoustic metrics are elaborated, which appear in the SU(N) symmetric Einstein-Yang-Mills-Higgs theory.

Now we assume that the kernel of the effective metric, ${\cal A}(\Theta,S,P^2)$, depends on three scalar arguments. The first one, $\Theta=\nabla_kU^k$, is the expansion scalar of the aether flow; in the Friedmann-type cosmology this scalar coincides with the triple Hubble function $\Theta=3H$. The second argument,
$S=\bar{\psi} \psi$ is the scalar describing the spinor particle number density; $\psi$ and $\bar{\psi}$ are the four-dimensional spinor and the Dirac conjugated spinor, respectively.  The third argument, $P=i \bar{\psi} \gamma^5\psi$, is the basic pseudoscalar; the covariant definition of the matrix $\gamma^5$ is discussed below.

The paper is organized as follows. In Section II, we reconstruct the extended Lagrangian of the model, and derive the modified master equations for the unit vector, spinor, axion and gravitational fields. The master equation for the guiding function $\Phi_*$ is obtained as a result of requirement of the invariance of the total Lagrangian with respect to the discrete transformations, associated with the basic axion field shift symmetry.
In Section III we consider the application of the established theory to the isotropic cosmological model, and derive the key equations for the axion field, for the gravity field and for the basic spinor invariants $S,P$. In Section IV we present and discuss exact solutions for three particular models. Section V includes discussion.

\section{The formalism}

\subsection{Action functional of the Einstein-Dirac-aether theory}

We use the action functional of the Einstein-Dirac-aether theory in the canonic form
\begin{equation}
{\cal S}_{(\rm EDA)} = \int d^4x\sqrt{-g}\left\{{-}\frac{1}{2\kappa}\left[R{+}2\Lambda {+} \lambda \left(g_{mn}U^m
U^n {-}1 \right) {+} K^{abmn} \nabla_a U_m  \nabla_b U_n  \right] \right. {+}
\label{EDA}
\end{equation}
$$
\left.
{+} \frac{i}{2}[\bar{\psi}\gamma^{k}D_{k}\psi {-}D_{k}\bar{\psi}\gamma^{k}\psi] {-} m \bar{\psi}\psi \right\}\,.
$$
The elements of this functional are well documented: $g$ is the determinant of the spacetime metric $g_{mn}$; $R$ is the Ricci
scalar; $\Lambda$ is the cosmological constant; $\kappa {=} 8\pi G$, $G$ is the Newtonian coupling constant, and $c{=}1$ in the chosen system of units. The four-vector $U^j$, is associated with the aether velocity; the term $\lambda \left(g_{mn}U^m U^n {-}1 \right)$ provides the four-vector  $U^j$ to be normalized to one;  $\lambda$ is the Lagrange multiplier. The kinetic term $K^{abmn} \ \nabla_a U_m \ \nabla_b U_n $ is quadratic in the covariant derivative
$\nabla_a U^m $ of the vector field $U^m$; the constitutive tensor $K^{abmn}$ includes the metric tensor $g^{ij}$ and the aether velocity four-vector $U^k$ only:
\begin{equation}
K^{abmn} {=} C_1 g^{ab} g^{mn} {+} C_2 g^{am}g^{bn}
{+} C_3 g^{an}g^{bm} {+} C_4 U^{a} U^{b}g^{mn}.
\label{2}
\end{equation}
The parameters $C_1$, $C_2$, $C_3$ and $C_4$ are the Jacobson's coupling constants \cite{J1}.

The spinor field with the seed  mass $m$ is described by two four-dimensional terms $\psi$ and $\bar{\psi}$ (the bar denotes the Dirac conjugation).
The symbol $\gamma^k$ relates to the Dirac matrices depending on coordinate; they are linked with the constant Dirac matrices in the Minkowski spacetime, $\gamma^{(a)}$, via the tetrad four-vectors $X^{m}_{(a)}$, which satisfy the relationships
\begin{equation}
g_{mn}X^{m}_{(a)}X^{n}_{(b)}=\eta_{(a)(b)} \,, \quad \eta^{(a)(b)}X^{m}_{(a)}X^{n}_{(b)}=g^{mn} \,,
\label{21}
\end{equation}
where $\eta_{(a)(b)}$ is the Minkowski metric. The matrices $\gamma^{k}=X^k_{(a)}\gamma^{(a)}$ satisfy the anti-commutation conditions
\begin{equation}
\gamma^m \gamma^n   {+}  \gamma^n \gamma^m =2 E g^{mn} \,,
\label{D7}
\end{equation}
where $E$ is the four-dimensional unit matrix.
Also, we use the matrix $\gamma^5$ defined as
\begin{equation}
\gamma^5 = -\frac{1}{4!} \epsilon_{mnpq} \gamma^m \gamma^n \gamma^p \gamma^q \,,
\label{gamma95}
\end{equation}
where $\epsilon_{mnpq}$ is the Levi-Civita tensor, which is connected with the completely anti-symmetric symbol $E_{mnpq}$ as follows:
\begin{equation}
\epsilon_{mnpq} = \sqrt{-g} E_{mnpq} \,, \quad E_{0123} = E_{(0)(1)(2)(3)} = -1 \,.
\label{Levi}
\end{equation}
Direct calculations show that
$$
\gamma^5 = -\frac{1}{4!} \epsilon_{mnpq} X^m_{(a)} X^n_{(b)} X^p_{(c)}X^q_{(d)} \gamma^{(a)} \gamma^{(b)} \gamma^{(c)} \gamma^{(d)} =
$$
\begin{equation}
{=} {-}\frac{1}{4!} \epsilon_{(a)(b)(c)(d)} \gamma^{(a)} \gamma^{(b)} \gamma^{(c)} \gamma^{(d)}  {=}
\gamma^{(0)} \gamma^{(1)}\gamma^{(2)}\gamma^{(3)} \equiv \gamma^{(5)} \,.
\label{gamma85}
\end{equation}
In other words, the matrix $\gamma^{5}=\gamma^{(5)}$ is a constant matrix in addition to the unit matrix $E$.
These details are important for further analysis, since in addition to the basic scalar  $S \equiv \bar{\psi}E \psi$, which describes the density of the spinor particles number, we use the basic pseudoscalar $P \equiv i \bar{\psi}\gamma^5 \psi$ (the imaginary unit $i$ in front  provides, that the matrix $i\gamma^5$ is real quantity).
The covariant derivatives of the spinor fields
\begin{equation}
D_{k}\psi = \partial_{k}\psi-\Gamma_{k}\psi \,, \quad
D_{k}\bar{\psi}=\partial_{k}\bar{\psi}+\bar{\psi}\Gamma_{k}
\label{D}
\end{equation}
contain the Fock-Ivanenko connection matrices $\Gamma_{k}$ defined as follows (see \cite{FI}):
\begin{equation}
\Gamma_{k}=\frac{1}{4}g_{mn}X^{(a)}_{s}\gamma^{s}\gamma^{n} \nabla_{k}X^{m}_{(a)}\,.
\label{5D}
\end{equation}

\subsection{Extended action functional of the axion field}

We assume that the action functional of the axion field is of the following form:
\begin{equation}
{\cal S}_{(\rm Axion)} = \int d^4x\sqrt{-g}\left\{ \frac12 \Psi_0^2 \left[G^{mn} \nabla_m \phi \nabla_n \phi - V(\phi, \Phi_*) \right] \right\} \,.
\label{Axion}
\end{equation}
As usual, the dimensionless pseudoscalar quantity $\phi$ describes the axion field, and the constant $\Psi_0$ is a quantity reciprocal to the constant of the axion-photon coupling $g_{A \gamma \gamma}$. The effective metric
\begin{equation}
G^{mn} = g^{mn} {+} {\cal A} U^m U^n
\label{A17}
\end{equation}
includes the kernel function ${\cal A} = {\cal A}(\Theta, S, P^2)$, which is the subject of the phenomenological modeling. If we assume additionally, that the effective metric is non-degenerated, and the inverse metric
$G_{jn} = g_{jn} {-} \frac{{\cal A}}{1+ {\cal A}} U_j U_n $ exists, we have to require that ${\cal A} \neq -1$.

The potential of the axion field is assumed to have the periodic form (see, e.g., \cite{Ax8} for details):
 \begin{equation}
V(\phi, \Phi_*) = \frac{m^2_A \Phi^2_*}{2 \pi^2} \left[1- \cos{\left(\frac{2 \pi \phi}{\Phi_*}\right)} \right] \,.
\label{A19}
\end{equation}
Here we introduced the so-called guiding function $\Phi_*(\Theta)$, where $\Theta = \nabla_k U^k$ is the expansion scalar of the  aether flow.
The potential is indicated as periodic, since the replacement $\phi \to \phi {+} k \Phi_*$  with integer $k$ leaves the potential unchanged.
The values of the axion field $\phi=n \Phi_*$ (with integer $n$) relate to the minima of the potential, i.e.,
\begin{equation}
V(n \Phi_*, \Phi_*) = 0 \,, \quad \left(\frac{\partial V}{\partial \phi}\right)_{| \phi {=} n \Phi_*} = 0 \,, \quad
\left(\frac{\partial^2 V}{\partial \phi^2}\right)_{| \phi {=} n \Phi_*} =   2 m^2_A  > 0 \,.
\label{A11}
\end{equation}

\subsection{Evolutionary equation for the guiding function $\Phi_*$}

We attract attention to the fact that the evolutionary equation for the guiding function $\Phi_*$ can be obtained from the requirement that the action functional (\ref{Axion}) is invariant with respect to the discrete transformation $\phi {=} \tilde{\phi} {+} k \Phi_*$.
The author of the review \cite{Ax8} uses for this case the terminology {\it discrete shift symmetry}, and, based on the formula (34) from \cite{Ax8}, one can assume that $\Phi_*$ is proportional to the axion vacuum expectation value, associated with the parameter $\frac{f_a}{\sqrt2}$ in \cite{Ax8}.
The periodic  potential of the axion field is invariant, i.e., $V(\phi, \Phi_*)=V(\tilde{\phi}, \Phi_*)$, and we have to require that the transformation of the kinetic part adds the complete divergence, which disappears due to the Gauss theorem. Indeed, after the discrete transformation we obtain that
\begin{equation}
{\cal S}_{(\rm Axion)} = \int d^4x\sqrt{-g}\left\{ \frac12 \Psi_0^2 \left[G^{mn} \nabla_m \left(\tilde{\phi} {+} k\Phi_* \right) \nabla_n \left(\tilde{\phi} {+} k\Phi_* \right) - V\left(\left(\tilde{\phi} {+}k \Phi_* \right), \Phi_*\right) \right] \right\} =
\label{A13}
\end{equation}
$$
= \tilde{{\cal S}}_{(\rm Axion)} + \frac12 k \Psi_0^2 \int d^4x\sqrt{-g} \ \nabla_m \left[G^{mn} \left(2\tilde{\phi}+
k \Phi_*\right) \nabla_n \Phi_* \right] -
$$
$$
- \frac12 k \Psi_0^2 \int d^4x\sqrt{-g} \  \left( 2\tilde{\phi} + k \Phi_*\right) \nabla_m \left(G^{mn} \nabla_n \Phi_*\right) \,.
$$
We omit the term with the complete divergence, and obtain that ${\cal S}_{(\rm Axion)}=\tilde{{\cal S}}_{(\rm Axion)}$, independently on the value of the variable $\tilde{\phi}$ and of the value of the integer $k$ , when
\begin{equation}
\nabla_m \left(G^{mn} \nabla_n \Phi_*\right) =0 \ \Rightarrow \ \frac{1}{\sqrt{-g}} \partial_m \left[\sqrt{-g} G^{mn} \partial_m \Phi_* \right] =0
\,.
\label{A14}
\end{equation}
Below we indicate the equation (\ref{A14}) as the evolutionary equation for the guiding function $\Phi_*$.

\subsection{Master equations}

\subsubsection{Extended master equation for the axion field }

Variation of the total action functional with respect to the pseudoscalar field $\phi$ gives the equation
\begin{equation}
\nabla_n \left[G^{mn} \nabla_m \phi \right] + \frac{m_A^2 \Phi_*}{2 \pi} \sin{\left(\frac{2 \pi \phi}{\Phi_*}\right)}
= 0 \,.
\label{ME1}
\end{equation}
Clearly, this equation is invariant with respect to the discrete transformation $\phi \to \phi + k \Phi_*$.
The evolutionary equation for the guiding function $\Phi_*$ (\ref{A14}) can be obtained from the equation (\ref{ME1}) if we put $\phi=\Phi_*$. In other words, $\Phi_*$ can be considered as the value of the axion field frozen into the first minimum of the axion potential.
Finally, the master equation (\ref{ME1}) can be rewritten in the form
\begin{equation}
\nabla_n \left(g^{mn} \nabla_m \phi \right) + \frac{m_A^2 \Phi_*}{2 \pi} \sin{\left(\frac{2 \pi \phi}{\Phi_*}\right)} = {\cal J} \,,
\label{ME11}
\end{equation}
where the pseudoscalar source in the right-hand size can be presented as follows:
\begin{equation}
{\cal J} = - \left[D {\cal A} + {\cal A}\left(\Theta + D \right) \right] D\phi \,, \quad D = U^k \nabla_k \,.
\label{ME111}
\end{equation}
It disappears, if the spinor kernel ${\cal A}$ vanishes.

\subsubsection{Master equations for the spinor field}

Variation with respect to $\bar{\psi}$ and $\psi$ gives the extended Dirac equations, which have the structure
\begin{equation}
i\gamma^{n} D_{n}\psi =  M \psi \,, \quad i D_{n}\bar\psi\gamma^{n} = -\bar{\psi} M  \,.
\label{D1}
\end{equation}
Now the matrix $M$ contains not only the canonic part $m E$, but also an additional (induced) part
\begin{equation}
M= m E  - \frac12 \Psi^2_0 {\dot{\phi}}^2 \left[E \frac{\partial {\cal A}}{\partial S} +i \gamma^5 \frac{\partial {\cal A}}{\partial P} \right] \,.
\label{D11}
\end{equation}
Here and below, for the sake of simplicity, we use the compact formula with $\dot{\phi} = D\phi = U^k \nabla_k \phi$:
\begin{equation}
M =  E \nu_1 + i \gamma^5 \nu_2 \,, \quad \nu_1 = m - \frac12 \Psi_0^2 {\dot{\phi}}^2 \frac{\partial {\cal A}}{\partial S} \,, \quad \nu_2 = - \frac12 \Psi_0^2 {\dot{\phi}}^2 \frac{\partial {\cal A}}{\partial P} \,.
\label{D111}
\end{equation}
Clearly, the modification of the Dirac equations is connected with the appearance of the axionically induced effective fermion mass
\begin{equation}
<M> \equiv \frac{\bar{\psi}M \psi}{{\bar\psi}\psi} = m - \frac12 \Psi^2_0 {\dot{\phi}}^2 \left[S \frac{\partial {\cal A}}{\partial S} +P \frac{\partial {\cal A}}{\partial P} \right] \,,
\label{D222}
\end{equation}
 which acquires an additional term proportional to the quantity ${\dot{\phi}}^2$. When the kernel ${\cal A}$ can be presented as ${\cal A}(\Theta,\frac{P^2}{S^2})$, the differential operator $S \frac{\partial {\cal A}}{\partial S} {+}P \frac{\partial {\cal A}}{\partial P} $ gives zero, and thus $<M> \equiv m$ independently on the state of the axion field.

\subsubsection{Master equations for the vector field}

Variation of the total functional with respect to the aether velocity four-vector gives the system of four equations
\begin{equation}
\nabla_a J^{a}_{\ j} = \lambda U_j + C_4 (DU_m)(\nabla_j U^m)- \kappa \Psi^2_0 \left\{{\cal A} D \phi \nabla_j \phi - \frac12 \nabla_j \left[(D\phi)^2 \frac{\partial {\cal A}}{\partial \Theta} \right] + \frac12 \nabla_j \left[\left(\frac{\partial V}{\partial \Phi_*}\right)  \ \left( \frac{\partial \Phi_*}{\partial \Theta}\right) \right]\right\} \,,
\label{D21}
\end{equation}
where the Jacobson tensor is of the canonic form
$J^{aj} {=} K^{abjn} \nabla_b U_n $.
The Lagrange multiplier can be found as
\begin{equation}
\lambda = U^j\nabla_a J^{a}_{\ j} - C_4 (DU_m)(DU^m)+ \kappa \Psi^2_0 \left\{{\cal A} (D \phi)^2 - \frac12 D \left[(D\phi)^2 \left(\frac{\partial {\cal A}}{\partial \Theta}\right) \right] + \frac12 D\left[\left(\frac{\partial V}{\partial \Phi_*}\right)  \  \left(\frac{\partial \Phi_*}{\partial \Theta}\right)\right]\right\}  \,.
\label{D221}
\end{equation}
Clearly, the term in (\ref{D21}) containing the kernel ${\cal A}$ can be considered as the source-term associated with the spinor extension of the Jacobson equations.

\subsubsection{Master equations for the gravity field}

Variation of the total action functional with respect to the metric gives the gravity field equations, which, formally speaking, have the canonic form
\begin{equation}
R_{pq}- \frac12 g_{pq} R  - \Lambda g_{pq} = T^{(\rm U)}_{pq} + \kappa T^{(\rm A)}_{pq} +  \kappa T^{(\rm D)}_{pq} \,.
\label{D2}
\end{equation}
However, since the Dirac matrices $\gamma^k$ and the tetrad four-vectors $X^j_{(a)}$ satisfy the relationships (\ref{D7}) and (\ref{21}),
respectively, they depend on the metric and thus, in the process of variation we have to use the following relationships:
\begin{equation}
\delta X^j_{(a)} = \frac14 \left[X_{p(a)}\delta^j_{q} + X_{q(a)}\delta^j_{p} \right]\delta g^{pq} \,, \quad \delta X_{j(a)} = - \frac14 \left[X_{p(a)}g_{jq} + X_{q(a)}g_{jp} \right]\delta g^{pq} \,,
\label{D764}
\end{equation}
\begin{equation}
\delta \gamma^{k} = \frac14 \delta g^{pq} \left(\gamma_p \delta^k_{q} + \gamma_q \delta^k_{p} \right) \,, \quad \delta \gamma_{k} = - \frac14 \delta g^{pq} \left(\gamma_p g_{kq} + \gamma_q g_{kp} \right) \,, \quad \delta \gamma^{(5)} = 0 = \delta \gamma^{5} \,.
\label{D769}
\end{equation}
The first source in the right-hand side of (\ref{D2})
$$
T^{(\rm U)}_{pq} =
\frac12 g_{pq} \ K^{abmn} \nabla_a U_m \nabla_b U_n{+} \lambda U_pU_q {+}
C_1\left[(\nabla_m U_p)(\nabla^m U_q) {-}
(\nabla_p U_m )(\nabla_q U^m) \right] {+}
$$
\begin{equation}
+C_4 {\cal D} U_p {\cal D} U_q {+}\nabla^m \left[U_{(p} J_{q)m} {-}
J_{m(p}U_{q)} {-}
J_{(pq)} U_m\right] \,,
\label{5Ein1}
\end{equation}
can be indicated as the stress-energy tensor of the vector field \cite{J1}. However, it is not a tensor, which contains the vector field only; since the Lagrange multiplier $\lambda$ is now of the form (\ref{D221}), it depends on the axion field via the terms ${\dot{\phi}}^2$ and $\Phi_*$, as well as, on the spinor field via the kernel ${\cal A}$ of the effective metric.
The term $T^{(\rm D)}_{pq}$ presents the canonic stress-energy tensor of the spinor field
\begin{equation}
T^{(\rm D)}_{pq} = - g_{pq} \left\{ \frac{i}{2} [\bar{\psi}\gamma^{k}D_{k}\psi {-}D_{k}\bar{\psi}\gamma^{k}\psi] {+} m \bar{\psi}\psi \right\}+
\frac{i}{4}\left[\bar\psi \gamma_{p} D_{q}\psi {+} \bar\psi \gamma_{q} D_{p}\psi {-} (D_{p}\bar\psi) \gamma_{q} \psi {-} (D_{p}\bar\psi) \gamma_{q} \psi \right] \,.
\label{D99}
\end{equation}
In contrast to the standard version the Lagrangian of the spinor field (the first term of this tensor with the metric in front) does not vanish on the solutions to the Dirac equations, since the effective mass $M \neq m E$ forms the sources in the right-hand sides of the equations (\ref{D1}).

The last source term describes the extended stress-energy tensor of the axion field
\begin{equation}
T^{(\rm A)}_{pq}= \Psi^2_0 \left\{\nabla_p \phi \nabla_q \phi  +
	\frac12 g_{pq}\left(V {-} G^{mn}\nabla_m \phi \nabla_n \phi  \right) - \frac12 \left[g_{pq}(D+\Theta) \left[(D\phi)^2 \left(\frac{\partial {\cal A}}{\partial \Theta}\right) - \left(\frac{\partial V}{\partial \Phi_*}\right) \left(\frac{\partial \Phi_*}{\partial \Theta}\right) \right] \right]\right\} \,.
\label{AX12}
\end{equation}
The total set of master equations is  derived, the next step is to apply this theory to the isotropic cosmological model.

\section{Cosmological application}

\subsection{Geometrical aspects of the model}

As an application we will consider the spatially isotropic homogeneous spacetime with the metric
\begin{equation}
ds^2 = dt^2 - a^2(t)[{dx^1}^2 + {dx^2}^2 + {dx^3}^2] \,,
\label{FLRW1}
\end{equation}
where $a(t)$ is the scale factor. We assume that all the fields inherits the spacetime symmetry and depend on the cosmological time only. The aether velocity four-vector has to be of the form $U^j = \delta^j_0$, and its covariant derivative is symmetric
\begin{equation}
\nabla_k U_m = H(t) \Delta_{km} \,, \quad  \Delta_{km}= g_{km} - U_k U_m \,.
\label{nablaU}
\end{equation}
The acceleration four-vector, shear tensor and vorticity tensor of the aether flow vanish, and the expansion scalar $\Theta {=} \nabla_k U^k$ is equal to the triple Hubble function $\Theta {=} 3H$, $H(t) \equiv \frac{\dot{a}}{a}$.

For the metric (\ref{FLRW1}) the tetrad four-vectors are of the form
\begin{equation}
X^i_{(0)} = U^i = \delta^i_0 \,, \quad X^i_{(\alpha)} = \delta^i_{\alpha} \frac{1}{a(t)} \,, \quad (\alpha = 1,2,3) \,,
\label{tetradF}
\end{equation}
the corresponding spinor connection coefficients are
\begin{equation}
\Gamma_0 = 0 \,, \quad \Gamma_{\alpha} = \frac12 \dot{a} \gamma^{(\alpha)} \gamma^{(0)} \,, \quad \gamma^{k} \Gamma_k = - \frac32 H \gamma^0 = -  \Gamma_k \gamma^{k} \,.
\label{GFriedmann}
\end{equation}
Also, we take into account that the sum of the Jacobson coupling constants $C_1{+}C_3$, estimated in 2017 as the result of observation of the binary neutron star merger (the events GW170817 and GRB 170817A \cite{170817}), is in the range $-6 \times 10^{-15}<C_1{+}C_3< 1.4 \times 10^{-15}$ (see, e.g., \cite{Dark} for details of estimations and references). In fact, we put below $C_3 {=} -C_1$. The parameter $C_4$ is hidden in this model, since the acceleration four-vector is vanishing, $DU^j=0$. This means that
the Jacobson tensor $J^{aj}$ is symmetric, and calculations of its nonzero components yield
\begin{equation}
 J^a_j = C_2 \Theta   \delta^a_j \,, \quad \nabla_a J^a_j = C_2 \nabla_j \Theta = 3 C_2 U_j \dot{H} \,.
\label{JFRiedmann}
\end{equation}
These results show that the equations (\ref{D21}) are satisfied, when $U^j=\delta^j_0$ and the Lagrange multiplier is of the form
\begin{equation}
\lambda = C_2 \dot{\Theta} + \kappa \Psi^2_0 \left\{{\cal A} {\dot{\phi}}^2  - \frac12 D \left[{\dot{\phi}}^2 \left(\frac{\partial {\cal A}}{\partial \Theta}\right) - \left(\frac{\partial V}{\partial \Phi_*}\right)  \  \left(\frac{\partial \Phi_*}{\partial \Theta}\right)\right]\right\} \,.
\label{FF1}
\end{equation}

\subsection{Reduced master equations}

\subsubsection{Key equation for the gravitational field}

The isotropic homogeneous cosmological model is so symmetric, that only two equations form the key system: first, the equation related to the choice $p=0$, $q=0$, second, the conservation law, as a consequence of the Bianchi identity. In our case the second equation is satisfied identically on the solutions to the axion, spinor and vector fields. Thus, the key equation for the gravity field is
\begin{equation}
3H^2 \Gamma -\Lambda =  mS +
\frac12 \kappa \Psi^2_0 \left\{(1+{\cal A}) \dot{\phi}^2 + V  + \Theta \left[\dot{\phi}^2 \frac{\partial {\cal A}}{\partial \Theta}  - \frac{\partial V}{\partial \Phi_*}  \  \frac{\partial \Phi_*}{\partial \Theta} \right] \right\} \,.
\label{keygrav}
\end{equation}
Here we introduced a new auxiliary parameter $\Gamma = 1 + \frac32 C_2 $.

\subsubsection{Reduced master equations for the spinor field and evolutionary equations of the spinor invariants}

We assume that the components of the spinor field are the functions of the cosmological time only; then the Dirac equations (\ref{D1}) yield
\begin{equation}
i \gamma^{0}\left(\partial_0 + \frac32 H \right) \psi = M \psi \,,
\quad i \left(\partial_0 + \frac32 H \right) \bar{\psi} \gamma^0 = - \bar{\psi} M \,.
\label{DF2}
\end{equation}
Using these equations we can find the rates of evolution of spinor invariant $S{=}\bar{\psi} \psi$, and two spinor pseudoinvariants: $P {=} \bar{\psi}i\gamma^5 \psi$ and $\Omega {=} \bar{\psi}\gamma^5 \gamma^0 \psi$.
First, we can find that
\begin{equation}
\frac{d}{dt}\left(\bar{\psi} \psi \right) = \frac{d}{dt}\left[\left(\bar{\psi} \gamma^0 \right) \left(\gamma^0 \psi \right)\right] = {-}3H \left(\bar{\psi} \psi \right) {+} i \bar{\psi} \left(M\gamma^0 {-} \gamma^0 M \right)\psi \,.
\label{dotS1}
\end{equation}
Using (\ref{D111}) one can  present the evolutionary equation for the scalar $S$ as follows:
\begin{equation}
\dot{S} + 3H S = - 2 \Omega \ \nu_2 \,.
\label{dotS2}
\end{equation}
Second, for the evolutionary equation for the pseudoinvariant $P$ we obtain
\begin{equation}
\dot{P} + 3H P = - \bar{\psi} \left(M\gamma^0 \gamma^5 {-} \gamma^5 \gamma^0 M \right)\psi \ \Rightarrow \dot{P} + 3H P =  2 \Omega \nu_1 \,.
\label{0dotP}
\end{equation}
Third, the evolutionary equation for the pseudoinvariant $\Omega$ is described by
\begin{equation}
\dot{\Omega} + 3H  \Omega = -i \bar{\psi} \left(M \gamma^5 + \gamma^5 M \right)\psi \ \Rightarrow \dot{\Omega} + 3H \Omega  = -2 P \nu_1 - 2 S \nu_2  \,.
\label{dotT}
\end{equation}
In other words, the set of functions $S(t)$, $P(t)$ and $\Omega(t)$ forms the closed evolutionary system.
Using these three evolutionary equations one can find that
\begin{equation}
\frac{d}{dt} \left(S^2-P^2-\Omega^2\right) + 6H \left(S^2-P^2 -\Omega^2\right) = 0 \,,
\label{dotT21}
\end{equation}
consequently, we deal with the first integral of this system:
\begin{equation}
a^6 \left(S^2-P^2-\Omega^2 \right) = {\rm const} \,.
\label{dotT22}
\end{equation}
It is convenient to introduce the variable $x= \frac{a(t)}{a(t_0)}$, the reduced dimensionless scale factor, where $t_0$ is some fixed moment of the cosmological time. If we introduce three auxiliary functions of this variable:
\begin{equation}
X = x^3 S \,, \quad Y = x^3 P \,, \quad Z = x^3 \Omega \,,
\label{dotT27}
\end{equation}
the first integral (\ref{dotT22}) takes the simplest form $X^2-Y^2-Z^2 = K$,
where $K$ is an arbitrary constant of integration.
In these terms the set of evolutionary equations can be rewritten as follows:
\begin{equation}
X^{\prime}(x) = - \frac{2}{x H} Z \nu_2 \,, \quad Y^{\prime}(x) =  \frac{2}{x H} Z \nu_1 \,, \quad Z^{\prime}(x) = -\frac{2}{xH} \left[\nu_1 Y + \nu_2 X \right] \,.
\label{X1}
\end{equation}

\subsubsection{Reduced equation for the axion field and for the guiding function}

The reduced equation (\ref{A14})
\begin{equation}
a^{-3} \frac{d}{dt}\left[a^3 (1+ {\cal A}) \frac{d \Phi_*}{dt} \right] =0
\label{X2}
\end{equation}
can be immediately integrated yielding
\begin{equation}
\dot{\Phi}_*(t) = \dot{\Phi}_*(t_0) \left(\frac{a(t_0)}{a(t)} \right)^3 \left(\frac{1+ {\cal A}(t_0)}{1+ {\cal A}(t)}\right) \,,
\label{X3}
\end{equation}
or equivalently, in terms of $x$
\begin{equation}
\Phi^{\prime}_*(x) = \frac{h}{x^4 H(x)(1+{\cal A}(x))} \,, \quad h= \dot{\Phi}_*(t_0) \left(1+ {\cal A}(t_0) \right)\,.
\label{X39}
\end{equation}

When we deal with the reduced equation (\ref{ME1})
\begin{equation}
a^{-3} \frac{d}{dt}\left[ a^3 (1+ {\cal A})\frac{d\phi}{dt} \right] + \frac{m_A^2 \Phi_*}{2 \pi} \sin{\left(\frac{2 \pi \phi}{\Phi_*}\right)}
= 0 \,,
\label{ME11}
\end{equation}
we see that (\ref{ME11}) converts into (\ref{X2}), if the axion field takes one of the equilibrium values $\phi=n \Phi_*$. This finding significantly simplifies the analysis of the formulated model and below we will use this ansatz.
Moreover, if the axions are in the equilibrium state, we obtain an additional simplification, since the quantity
\begin{equation}
\frac{\partial V}{\partial \Phi_*} = \frac{m_A^2 \Phi_*}{\pi^2} \left[1-\cos{\left(\frac{2\pi \phi}{\Phi_*}\right)} - \frac{\pi \phi}{\Phi_*} \sin{\left(\frac{2\pi \phi}{\Phi_*}\right) } \right]
\label{ME191}
\end{equation}
happens to be equal to zero in this case, and the key equation for the gravity field (\ref{keygrav}) takes the form
\begin{equation}
3H^2 \Gamma - \Lambda =  mS +
\frac12 \kappa n^2 \Psi^2_0 \left[\dot{\Phi}_*(t_0) \left(\frac{a(t_0)}{a(t)} \right)^3 \left(1+ {\cal A}(t_0)\right)\right]^2 \left(1- H \frac{\partial}{\partial H} \right) \left( \frac{1}{1+{\cal A}} \right) \,.
\label{ME181}
\end{equation}
Thus, we obtained a model, for which a simple receipt of its study exists. First, one has to formulate a physically motivated kernel of the aetheric effective metric, i.e., the function ${\cal A}(H,S,P^2)$. Second, from the equations (\ref{dotS2}) - (\ref{dotT}) one has to find the functions $S$ and $P$, and obtain the information concerning the evolution of the spinor particle number density ${\cal N}= S$. Third, one has to solve the equation (\ref{ME181}) with found functions $S$ and $P$, thus obtaining the Hubble function $H$, scale factor $a$ and acceleration parameter $-q$. Fourth, one has to find the guiding function $\Phi_*$ thus describing the evolution of the basic states of the axion system.
Below we used this receipt and analyzed three exactly integrable models.

\section{Three examples of exactly integrable models}

\subsection{First example}

For the first model we have chosen the function ${\cal A}$ so that
\begin{equation}
{\cal A} = \frac{1}{H F(S)} -1 \ \Rightarrow \frac{1}{1+{\cal A}} = H F(S) \,.
\label{1M1}
\end{equation}
Since in this first model the kernel ${\cal A}$ does not depend on the argument $P$, we see that $\nu_2=0$ and thus the solution to (\ref{dotS2}) is
$S(x) = \frac{S(t_0)}{x^3}$. Since $\left(1- H \frac{\partial}{\partial H} \right) H F(S)=0$, we obtain the key equation for the gravity field in the form
\begin{equation}
3H^2 \Gamma - \Lambda = m S(t_0) x^{-3} \,.
\label{1M2}
\end{equation}
It is important to emphasize that only the spinor field forms the source in the right-hand size of this equation. The contribution of the axion field is switched off due to the specific guidance of the axion field evolution, which is performing by the spinor field via the effective aetheric metric.
The equation (\ref{1M2}) gives the known solution for the scale factor in the form
\begin{equation}
\frac{a(t)}{a(t_0)} = \left\{\cosh{\left[\frac32 H_{\infty}(t{-}t_0)\right]} {+} \sqrt{1{+}\rho_0^2} \ \sinh{\left[\frac32 H_{\infty}(t{-}t_0)\right]} \right\}^{\frac23} \,.
\label{1M3}
\end{equation}
Here we used the following notations:
\begin{equation}
H_{\infty} = \sqrt{\frac{\Lambda}{3\Gamma}} \,, \quad \rho_0 = \sqrt{\frac{m S(t_0)}{\Lambda}} \,.
\label{1M4}
\end{equation}
In terms of cosmological time the Hubble function takes the form
\begin{equation}
\frac{H(t)}{H_{\infty}} = \frac{\tanh{\left[\frac32 H_{\infty}(t{-}t_0)\right]} + \sqrt{1+\rho_0^2}}{1+ \sqrt{1+\rho_0^2}\tanh{\left[\frac32 H_{\infty}(t{-}t_0)\right]}} \,, \quad H(t_0) = H_{\infty} \sqrt{1+\rho^2_0} \,.
\label{1M5}
\end{equation}
Clearly, in the asymptotic regime $H(t \to \infty) \to H_{\infty}$, and $a(t \to \infty) \propto e^{H_{\infty}t}$.
The acceleration parameter
\begin{equation}
-q(t) = 1- \frac{\frac32 \rho_0^2}{\left\{\sinh{\left[\frac32 H_{\infty}(t{-}t_0)\right]} {+} \sqrt{1{+}\rho_0^2} \cosh{\left[\frac32 H_{\infty}(t{-}t_0)\right]} \right\}^2}
\label{1M6}
\end{equation}
tends asymptotically to one, $-q(\infty) =1$, and has the initial value $-q(t_0) = \frac{2-\rho_0^2}{2(1+\rho_0^2)} $.
When $\rho_0 > \sqrt2$, i.e., $m S(t_0)>2\Lambda$, the acceleration parameter changes the sign, thus indicating that the deceleration epoch is changing by the acceleration one at the moment
\begin{equation}
t=t_* = t_0+ \frac{2}{3H_{\infty}} {\rm Arsinh}{\left(\frac{\sqrt{1+\rho^2_0}-\sqrt3}{\sqrt2 \rho_0} \right)}.
\label{1M69}
\end{equation}
In order to find the guiding function, we specify the function $F(S)$ as follows:
\begin{equation}
F(S) = \frac{F_0}{1+ \left(\frac{S(t)}{S(t_0)}\right)^2} \Rightarrow F(S=0) = F_0 \neq 0 \,, \quad F(S(t_0)) =\frac12 F_0 \,.
\label{1M8}
\end{equation}
In this case the parameter $h$ takes the value
\begin{equation}
h = \dot{\Phi}_*(t_0)(1+{\cal A}(t_0)) = \frac{2}{F_0 H(t_0)} \dot{\Phi}_*(t_0) \,,
\label{1M95}
\end{equation}
and the derivative of the guiding function takes the form
\begin{equation}
\Phi^{\prime}_*(x) = \left[\frac{2\dot{\Phi}_*(t_0)}{H_{\infty} \sqrt{1+\rho^2_0}} \right] \frac{ x^2}{\left(1+ x^6 \right)} \,.
\label{1M9}
\end{equation}
Final integration gives
\begin{equation}
\Phi_*(t) = \Phi_*(t_0) +  \frac{2\dot{\Phi}_*(t_0)}{3H_{\infty}\sqrt{1+\rho^2_0}}  \left\{\arctan{\left[\cosh{\left(\frac32 H_{\infty}(t{-}t_0)\right)} {+} \sqrt{1{+}\rho_0^2} \sinh{\left(\frac32 H_{\infty}(t{-}t_0)\right)}  \right]^2}-\frac{\pi}{4} \right\} \,.
\label{1M11}
\end{equation}
Since we started our analysis using the ansatz $\Phi_*{=}\Phi_*(\Theta)$, and since $\Theta{=}3H$, the formula (\ref{1M11}) can be rewritten as
\begin{equation}
\Phi_*(H) = \Phi_*(H(t_0)) +  \frac{2\dot{\Phi}_*(t_0)}{3H_{\infty}\sqrt{1+\rho^2_0}}  \left\{\arctan{\left[\frac{\rho^2_0 H^2_{\infty}}{H^2-H^2_{\infty}} \right]}-\frac{\pi}{4} \right\} \,.
\label{2M11}
\end{equation}
In the asymptotic regime the guiding function tends to constant
\begin{equation}
\Phi_*(t \to \infty) \to \Phi_*(t_0) + \frac{\pi \dot{\Phi}_*(t_0)}{6 H_{\infty}\sqrt{1+\rho^2_0}}  \,.
\label{1M131}
\end{equation}
The function $\Phi_*(t)$ monotonically grows, when $\dot{\Phi}_*(t_0)$ is positive; in principle,                                                                                                                                                                                                                                                                                                                                                                                                                                                                                                                                                                                                                                                                                                                                                                                                                                                                                                                                                                                                                                                                                                                                                                                                                                                                                                                                                                                                                                                                                                                                                                                                                                                                                                                                                                                                                                                                                                                                                                                                                                                                                                                                                                                                                                                                                                                                                                                                                                                                                                    varying the value of the parameter $\dot{\Phi}_*(t_0)$, one can obtain that $\Phi_*(\infty) >> \Phi_*(t_0)$, i.e., an asymptotic growth of the axion field takes place.

\subsection{Second exactly integrable model}

As the second illustration, we consider the model, in which the kernel ${\cal A}$ does not depend on $S$, but is the function of the square of spinor pseudoinvariant, i.e., ${\cal A}={\cal A}(\Theta, P^2)$. Also, we assume that the seed mass is equal to zero, $m=0$. In this case one obtains that $\nu_1=0$ and thus the equation (\ref{0dotP}) gives the solution $P(x)=\frac{P(t_0)}{x^3}$. Our choice for the kernel of the effective metric is
\begin{equation}
\frac{1}{1+{\cal A}} = \frac{\Theta}{3} F_1 \log{\left[\frac{P^2_* + P^{2}(t)}{P^2(t_0)}\right]} \,,
\label{1P22}
\end{equation}
where $P_*$ is some phenomenological parameter with the dimensionality of $P$, and $F_1$ some constant parameter.
Since $m=0$, the key equation (\ref{1M2}) for the gravity field gives the solution of the de Sitter type
\begin{equation}
H(t) = H_{\infty} \,, \quad a(t) =a(t_0) e^{H_{\infty}(t-t_0)} \,.
\label{dS1}
\end{equation}
Then it is convenient to introduce the dimensionless parameter $x^3_* {=} \frac{P(t_0)}{P_*}$, and to rewrite the first and third equations (\ref{X1}) in the form
\begin{equation}
\frac{dX}{d\xi} = - Q Z \,, \quad \frac{dZ}{d\xi} = - Q X \,,
\label{1P23}
\end{equation}
where we used the new convenient variable $\xi$ and auxiliary parameter $Q$:
\begin{equation}
\xi = \arctan{\left(\frac{x}{x_*} \right)^3} \,, \quad Q = \frac{2 n^2 \Psi^2_0 {\dot{\Phi}_*}^2(t_0) F_1}{3 P_*} \,.
\label{1P24}
\end{equation}
For the function $X$, which we search for, the key equation is $X^{\prime \prime}(\xi) = Q^2 X $
and we obtain the solution
\begin{equation}
X(x) = x^3 S(x) = S(1) \cosh{[Q u(x)]} + \left[S^{\prime}(1)+ 3S(1) \right] \frac{(1+x^6_*)}{3x^2_*} \sinh{[Q u(x)]} \,,
\label{1P27}
\end{equation}
where the argument $u(x)$, which satisfies null initial condition $u(1)=0$, is
\begin{equation}
u(x)= \arctan{\left(\frac{x}{x_*} \right)^3}{-}\arctan{\left(\frac{1}{x_*} \right)^3} \,.
\label{1P28}
\end{equation}
Asymptotic behavior of the obtained function is the following:
\begin{equation}
u(\infty)= \frac{\pi}{2} {-}\arctan{\left(\frac{1}{x_*} \right)^3} \,, \quad S(\infty)=0 \,.
\label{1P31}
\end{equation}
In order to illustrate the behavior of the function $S$, which describes the spinor particle number density ${\cal N}{=}S$, we simplify the general situation by two conditions: first, $S(1)=0$, i.e., the initial value is vanishing; second, $x_*=1$, i.e., $P_*=P(t_0)$. Then the function
\begin{equation}
S(x) = \frac23 S^{\prime}(1) x^{-3} \sinh{[Q u(x)]}
\label{1P27}
\end{equation}
starts and finishes with equal values $S(1){=}0{=}S(\infty)$. This means that  the function $S(x)$ has at least one extremum. Analyzing the derivative
of this function
\begin{equation}
S^{\prime}(x) = \frac{2 S^{\prime}(1)}{x^{4}} \left[ -\sinh{[Q u(x)]} + \left(\frac{Qx^3}{1+x^6}\right) \cosh{[Q u(x)]} \right] \,,
\label{1P297}
\end{equation}
we see that on the finite interval of $x$ it takes zero value only once, namely at $x=x_{\rm m}$, which is the solution to the transcendent equation
\begin{equation}
\frac{1}{Q}\tanh{[Q u(x)]} = \frac{x^3}{1+x^6}  \,.
\label{3P297}
\end{equation}
Indeed, the function in the left-hand side of this equation starts from zero value and grows monotonically; the function in the right-hand side starts from the value $\frac12$ and decreases monotonically. For arbitrary $Q$ we can find only one point of intersection of the corresponding graphs, this point  relates to the maximum of the function $S(x)$.
In other words, the spinor particle number density starts from zero value, grows exponentially up to maximal value $S_{\rm max}$ (see (\ref{1P27}) at $x=x_{\rm m}$), and then decreases monotonically, reflecting the law of Universe expansion. The growth of the function $S$ can be interpret as the effect of the Universe spinorization, and the maximal value $S(t_{\rm m})$ relates to the number of relic spinor particles created in the early Universe.

In the case under consideration the equation for the guiding function has the form
\begin{equation}
\Phi^{\prime}_*(x) = h_1 x^{-4}\log{\left(\frac{1}{x^6_*} + \frac{1}{x^6} \right)} \,, \quad h_1 = \frac{\dot{\Phi}_*(t_0)}{H_{\infty} \log{(1+x^{-6}_*)}} \,.
\label{P77}
\end{equation}
The solution to this equation is
\begin{equation}
\Phi_*(x) = \Phi_*(t_0) {-} \frac{1}{3}h_1 \left\{
\left(1{-}\frac{1}{x^3}\right)(2 {+} \log{x_*^6}) {+}
\frac{1}{x^3}\log{\left(1{+}\frac{x^6_*}{x^6} \right)} {-} \log{(1{+}x^6_*)}
{+} \frac{2}{x_*^3}\left[\arctan\left(\frac{x_*}{x}\right)^3 {-} \arctan{x_*^3}\right]
\right\} \,.
\label{P777}
\end{equation}
The function $\Phi_*(x)$ is monotonic, and the asymptotic value of this function is
\begin{equation}
\Phi_*(\infty) = \Phi_*(t_0) {-} \frac{1}{3}h_1 \left\{
2 {+} \log{x_*^6} {-} \log{(1{+}x^6_*)}
{-} \frac{2}{x_*^3}\arctan{x_*^3} \right\} \,.
\label{P787}
\end{equation}
For illustration, we can consider $x_*=1$ and obtain that
\begin{equation}
\Phi_*(\infty) - \Phi_*(t_0) \approx   0.125 \frac{\dot{\Phi}_*(t_0)}{H_{\infty}}  \,.
\label{P787}
\end{equation}
Clearly,  the guiding function monotonically grows or decreases, depending on the sign of the initial parameter $\dot{\Phi}_*(t_0)$.

\subsection{Third exactly integrable model}

In the third illustration of the model we consider again the case ${\cal A}(\Theta,S)$ with $F(S)$ given by (\ref{1M8}), but now the function
$\frac{1}{1+{\cal A}} = \frac{\Theta^2}{9} F(S)$ is quadratic in $\Theta$.
As in the first model, we obtain the solution $S(x)=S(t_0)x^{-3}$, but the key equation for the gravity field (\ref{ME181}) takes more complicated form
\begin{equation}
H^2 = \frac{\Lambda}{3 \Gamma} + \frac{mS(t_0)}{3 \Gamma x^3} -
\frac{2 \kappa n^2 \Psi^2_0 {\dot{\Phi}}^2_*(t_0)}{3 \Gamma F_0 H^4(t_0)} \left( \frac{H^2}{1+x^6}\right) \,.
\label{ME981}
\end{equation}
The positive solution for the Hubble function is
\begin{equation}
H(x)  = H_{\infty} \sqrt{\left(1 +  \frac{\rho^2_0}{x^3} \right) \left( \frac{1+x^6}{B + x^6}\right)} \,,
\label{ME981}
\end{equation}
where the following notations are introduced:
\begin{equation}
B=1 + \frac{2 \kappa n^2 \Psi^2_0 {\dot{\Phi}}^2_*(t_0)}{3 \Gamma F_0 H^4(t_0)} \,, \quad \rho_0^2=\frac{mS(t_0)}{\Lambda} \,, \quad H(t_0)=  H_{\infty} \sqrt{ \frac{2\left(1 + \rho^2_0 \right)}{B +1}} \,.
\label{ME981}
\end{equation}
The acceleration parameter is
\begin{equation}
-q(x) = 1 - \frac{3\rho^2_0}{2(x^3+\rho^2_0)} + \frac{3(B-1)x^6}{(1+x^6)(B+x^6)} \,, \quad -q(\infty) =1 \,.
\label{ME098}
\end{equation}
In general case, the corresponding scale factor $a(t)$ can be found only in quadratures, however, when $m=0$, the solution can be presented in the following implicit form
\begin{equation}
\left|\frac{\sqrt{B(1+x^6)}-\sqrt{B+x^6}}{\sqrt{B(1+x^6)}+\sqrt{B+x^6}} \right|^{\sqrt{B}}\left|\frac{\sqrt{1+x^6}+\sqrt{B+x^6}}{\sqrt{B+x^6}-\sqrt{1+x^6}} \right| =
\label{M190}
\end{equation}
$$
\left|\frac{\sqrt{2B}-\sqrt{B+1}}{\sqrt{2B}+\sqrt{B+1}} \right|^{\sqrt{B}}\left|\frac{\sqrt{2}+\sqrt{B+1}}{\sqrt{B+1}-\sqrt{2}} \right|
\times \exp{\left[6H_{\infty}(t-t_0)\right]} \,.
$$
Searching for the guiding function $\Phi_*(x)$, we obtain in this case the integral
\begin{equation}
\Phi_*(x) = \Phi_*(t_0) + \frac{2 \dot{\Phi}_*(t_0)H_{\infty}}{H^2(t_0)} \int_1^x \frac{z^2 dz}{\sqrt{(B + z^6)(1+z^6)}} \,.
\label{Mjk0}
\end{equation}
Again the guiding function grows or decreases monotonically depending on the value of the parameter  $\dot{\Phi}_*(t_0)$. The integral in (\ref{Mjk0}) is regular at infinity and predetermines the asymptotic value $\Phi_*(\infty)$ (as the function of the parameter $B$). A typical behavior of the functions $a(t)$ and $\Phi_*(x(t))$ can be illustrated, e.g., by the particular case $B=0$, which relates to the following choice of the parameter $F_0$:
$F_0= - \frac{2 \kappa n^2 \Psi^2_0 {\dot{\Phi}}^2_*(t_0)}{3 \Gamma H^4(t_0)}<0$.
For this particular case we obtain
\begin{equation}
\frac{a(t)}{a(t_0)} = x(t)= \left\{\cosh{\left[3H_{\infty}(t-t_0) \right]} + \sqrt2 \sinh{\left[3H_{\infty}(t-t_0) \right]} \right\}^{\frac13} \,,
\label{M2309}
\end{equation}
\begin{equation}
H(x) = H_{\infty} \sqrt{1+ x^{-6}} \,, \quad H(t) = H_{\infty} \left\{\frac{\tanh{\left[3H_{\infty}(t-t_0)\right] + \sqrt2 }}{1+ \sqrt2 \tanh{\left[3H_{\infty}(t-t_0) \right]}} \right\}\,, \quad  H(t_0) = H_{\infty} \sqrt{2}  \,,
\label{M2709}
\end{equation}
\begin{equation}
\Phi_*(x) = \Phi_*(t_0) + \frac{\dot{\Phi}_*(t_0)}{6H_{\infty}} \log{\left[\frac{\left(\sqrt{1+x^6}-1 \right)(\sqrt2 +1)}{\left(\sqrt{1+x^6}+1 \right)(\sqrt2 -1)} \right]} \,, \quad \Phi_*(\infty) = \Phi_*(t_0) + \frac{\dot{\Phi}_*(t_0)}{3H_{\infty}} \log{(\sqrt2 +1)} \,.
\label{M230}
\end{equation}
In terms of $H$ the guiding function takes the form
\begin{equation}
\Phi_*(H) = \Phi_*(H(t_0)) + \frac{\dot{\Phi}_*(t_0)}{6H_{\infty}} \log{\left[\frac{(\sqrt2 +1)\left(H-\sqrt{H^2-H^2_{\infty}}\right)}{(\sqrt2 -1)\left(H+\sqrt{H^2+H^2_{\infty}}\right)} \right]} \,.
\label{M2770}
\end{equation}
 This function predetermines the current level of basic equilibrium states $\phi(t)=n \Phi_*(t) $ of the axions, which form the axionic part of the cosmic dark matter.

\section{Discussion}

The presented model is quite simple, but it shows unexpectedly interesting results. Indeed, we have modified the Lagrangian of the axion field only by introducing the effective metric $G^{mn} {=} g^{mn}{+}{\cal A}(\Theta,S,P^2)U^m U^n$ instead of spacetime metric $g^{mn}$, and can find the following three details.

Since the kernel of this extended  metric, ${\cal A}$, contains the spinor field, which is encoded in the scalar $S$ and psedoscalar $P$, we obtain immediately the modification of the effective mass matrix $M$ (\ref{D11}) and the corresponding complication of the evolutionary equations for the quantities $S$ and $P$ (see, (\ref{dotS2})-(\ref{dotT22})). Modifications of similar type are obtained in the works of Saha (see, e.g., \cite{Saha1,Saha2}), but the corresponding modifications were included into the Lagrangian of the spinor field itself, and they are not associated, formally speaking, with the axion field. In the presented version, the effective mass $<M>$ contains the term proportional to the square of the derivative of the field ${\dot{\phi}}^2$ (see (\ref{D222})). In other words, oscillations of the axion field induce directly the variations of the spinor particle effective mass; especially, it is interesting, when we deal with the neutrinos with vanishing seed mass, $m=0$ \cite{BEfirst}. However, the effect of axionically induced effective mass disappears, when the quantities $S$ and $P^2$ enters the kernel in the so-called self-similar combination $\frac{P^2}{S^2}$.

Since the extended  metric depends on the aether velocity four-vector $U^k$ via the expansion scalar of the aether flow, $\Theta {=} \nabla_k U^k$, and directly via the term $U^m U^n$, we obtain modifications of the Jacobson equations (see (\ref{D21}), (\ref{D221})), which again contain the term ${\dot{\phi}}^2$ associated with the rate of the axion field evolution.

Since the extended  metric depends on the spacetime metric via the expansion scalar $\Theta {=} \nabla_k U^k$, we find additional terms in the right-hand sides of the gravity field equations (see $\lambda$ in (\ref{5Ein1}) and the term with $g_{pq}$ in front in (\ref{AX12})). The most interesting detail of these source terms appears in the formula (\ref{ME181}), which contains the term $\left(1{-} \Theta \frac{\partial}{\partial \Theta} \right) \left( \frac{1}{1{+}{\cal A}} \right)$. If the function $\frac{1}{1{+}{\cal A}}$ is linear in the expansion scalar $\Theta {=}3H$, the mentioned term vanishes and thus the information about the axion field disappears from the key equation for the gravity field, or in other words, the axion field becomes hidden from the point of view of gravity.

The guiding function $\Phi_*(\Theta)$, which enters the total Lagrangian via the axion field potential $V(\phi,\Phi_*)$ (see (\ref{A19})), attracts special attention. In the canonic axion field theory this function is considered as a constant, $\Phi_*=2\pi f_a$, so that the discrete transformation $\phi \to \phi + 2\pi f_a k$ leaves the Lagrangian invariant ($k$ is an integer). The operator of differentiation also leaves the kinetic part of the Lagrangian invariant, and the canonic term $\phi F^*_{mn}F^{mn}$ is invariant, since $2\pi n F^*_{mn}F^{mn}$ can be rewritten as a complete divergence ($F_{mn}$ is the Maxwell tensor, $F^*_{pq}$ is its dual pseudotensor, see, e.g., \cite{Ax1}-\cite{Ax4}). In our case $\Phi_*(\Theta)$ depends on coordinates via the scalar $\Theta$. When we consider the discrete transformation $\phi \to \phi + k\Phi_*(\Theta)$, the potential does not change its value. However, $\nabla_k \Phi_*$ is not now vanishing quantity, and we have to formulate the requirement, which admits us to speak about the invariance with respect to mentioned discrete transformation. We focused on this problem in the Subsection IIC and have shown that, the extended Lagrangian of the axion field is invariant with respect to the discrete transformation $\phi \to \phi + k \Phi_* $, if the guiding function satisfies the equation (\ref{A14}). This equation coincides with the master equation for the axion field, when the value $\phi$ relates to some minimum of the axion field potential, $\phi = n \Phi_*$. This fact allows us to indicate the master equation for the guiding function $\Phi_*$ as the equation describing the basic equilibrium state of the axion field. In the papers \cite{Dark,BS1} we considered this statement as an {\it ansatz}. Now (this is a reason to thank the anonymous reviewer of the article \cite{BS1}), we can state that master equation (\ref{A14}) for the guiding function $\Phi_*$ is not just an ansatz, but a mathematically proven statement.
The formulas (\ref{2M11}) and (\ref{M2770}), which present the exact solutions to (\ref{A14}) for two particular exactly integrable models, illustrate the following statement: the guiding function $\Phi_*$ can be chosen as regular monotonic function of the cosmological time, which tends asymptotically to some finite value $\Phi_*(\infty)$, which can be associated with the axion vacuum expectation value for the late-time Universe evolution stage.

Finally, we would like to attract attention to the backreaction of the spinor field evolution on the cosmodynamics. First, using three examples of the exactly integrable models, we illustrated the idea that the choice of the spinor kernel of the effective metric, ${\cal A}(\Theta,S,P^2)$, has a significant impact on the behavior of the Hubble function, scale factor and acceleration parameter. Second, this choice predetermines the behavior of the guiding function $\Phi_*(t)$ and its asymptotic value $\Phi_*(\infty)$, which is connected with the axion vacuum expectation value at the late-time period of the Universe evolution.

\end{document}